\documentstyle[11pt,aussois2003,twoside,psfig,epsfig]{article}
\def\simlt{\lower.5ex\hbox{$\; \buildrel < \over \sim \;$}}
\def\simgt{\lower.5ex\hbox{$\; \buildrel > \over \sim \;$}}
\def\etal{{\it et al.}\ }

\def\edcomment#1{\iffalse\marginpar{\raggedright\sl#1\/}\else\relax\fi}
\reversemarginpar
\begin{document}
%
\title{Probing Dark Matter and Dark Energy with Space-Based Weak Lensing}
%
\author{Richard Massey}
\affil{Institute of Astronomy, Madingley Road, Cambridge CB3 OHA, UK}
\author{Alexandre Refregier}
\affil{Service d'Astrophysique, CEA/Saclay, 91191 Gif sur Yvette, France}
\author{Jason Rhodes}
\affil{California Institute of Technology, 1201 E. California Blvd.,
Pasadena, CA 91125, USA}


\label{page:first}

\begin{abstract}

 Weak lensing provides a direct measure of the distribution of mass in the
 universe, and is therefore a uniquely powerful probe of dark matter. Weak
 lensing can also be used to measure the twin phenomenon of dark energy, via
 its effect upon the cosmological growth rate of structures. Essential for this
 technique are well-resolved images of background galaxies out to large
 distances. As a concrete example of the surveys that will become available by
 the end of the decade, we consider the planned {\it Supernova/Acceleration
 Probe} (SNAP) space telescope. Detailed simulations of space-based images,
 manufactured using the shapelets formalism, enable us to quantitatively
 predict the future sensitivity to weak lensing shear. The high number density
 of galaxies resolved from space will enable maps of dark matter to be produced
 in two and three dimensions, with a resolution superior to that from the
 ground. Such observations will also afford reduced systematics for
 high-precision measurements of weak lensing statistics. These will be used to
 set tight constraints on cosmological parameters. In particular, the parameter
 for equation of state of dark energy, $w$, will be measured using weak lensing
 with a precision comparable to and somewhat orthogonal to constraints from
 other methods.

\end{abstract}

\section{Introduction} \label{intro} 

During its journey to our telescopes, the light from background galaxies is
deflected by foreground mass concentrations, which act as gravitational lenses
along the line-of-sight. The observed distortions in the shapes of distant
galaxies are directly related to the gravitational potential of foreground
structures, and therefore to their mass. This effect is independent of the
nature or state of the foreground mass and therefore traces the distribution of
otherwise invisible dark matter. In recent years, many groups have measured
coherent distortions in the shapes of galaxies to measure the mass distribution
of clusters ({\it e.g.}~Joffre {\it et al.}~2000; Clowe, Trentham \& Tonry
2001; Dahle {\it et al.}~2002). Other groups have measured the distribution of
large-scale structure in random regions in the sky to set constraints on
cosmological parameters ({\it e.g.}~van Waerbeke {\it et al.}~2002; Bacon {\it
et al.}~2002; Hoekstra {\it et al.}~2002; Jarvis {\it et al.}~2003). The second
moment of the resulting ``cosmic shear'' field already provides a constraint on
the amplitude of the mass power spectrum $\Omega_m \sigma_8^{0.5}$ which is
comparable to the constraints set by more traditional methods such as the
abundance of x-ray selected clusters ({\it e.g.}~Viana \& Liddle~1999;
Pierpaoli, Scott \& White~2001). The higher order moments of the shear field
promise to break the $\Omega_m$-$\sigma_8$ degeneracy (Bernardeau et al.
\etal1997) in the next generation of large cosmic shear surveys, {\it
e.g.}~CFHT Legacy Survey (CFHTLS, Mellier \etal2000), LSST (Tyson \etal 2002a)
and Pan-STARRS (Kaiser, Tonry \& Luppino 2000). For reviews of the techniques
and current status of weak lensing measurements, other than this volume, see
Mellier et al. (2001), Hoekstra et al. (2002), Wittman (2002) and Refregier
(2003).

The rapid growth of dedicated surveys and improvements in shear measurement
methods has meant that errors on cosmological parameters from cosmic shear are
now limited by systematic rather than statistical errors. Weak lensing has the
advantage that its systematics arise from imperfect instruments and image
analysis, rather than unknown physics (such as the mass-temperature relation
which dominates $x$-ray selected cluster samples (Huterer \& White, 2002)).
Image analysis tools are currently making important advances, from the
well-tested method of Kaiser, Squires \& Broadhurst (1995; KSB), with work by
Rhodes, Refregier \& Groth (2000; RRG), Bernstein \& Jarvis (2003) and
``shapelets'' by Refregier (2003a). Meeting the engineering challenge, however,
will inevitably require rising above the atmosphere. Convolution with a large
Point-Spread Function (PSF) that is larger than the majority of distant (and
therefore small) objects irretrievably destroys any information their shapes
had contained. This limits the number density and redshift of galaxies that can
be used to measure cosmic shear with ground-based lensing surveys. Convolution
with the time-varying atmospheric PSF further limits the recovery of the shape
information of the remaining galaxies to the accuracy with which the PSF can be
modelled from one exposure to the next.

In these proceedings, we discuss the potential advances that a
wide-field imager from space offers for weak lensing. Adopting
the planned {\it Supernova/ Acceleration Probe} (SNAP) mission as a
concrete example, we present detailed shapelet-based simulations
used to estimate the lensing sensitivity of such observations. Most
interestingly for cosmology, SNAP's enhanced spatial resolution will
capture the shapes of background galaxies up to a much greater
distance. Space-based surveys will therefore probe the evolution
of the mass distribution, and the growth of structures, over a large
fraction of time in the evolution of the universe. This evolution
provides strong constraints on both $\Omega_m$ and the equation
of state parameter of dark energy $w$.  As discovered from
observations of type Ia superov\ae~({\it e.g.}~Perlmutter \etal1999),
this dark energy or ``quintessence'' is accelerating the expansion of
the universe and hence retarding the growth of the mass power
spectrum. Earlier studies of the constraints on dark energy from
weak lensing surveys can be found in Hu (1999), Huterer (2001),
Benabed \& Bernardeau (2001), Hu (2001), Weinberg \& Kamionkowski
(2002), Munshi \& Wang (2002), Benabed \& van Waerbeke (2003).
We also compare the quality of dark matter maps that will be
possible from SNAP with those possible from current ground-based
surveys. High resolution maps from space will enable the production of
a fully mass-selected cluster catalogue, useful for a further
constraint on cosmological parameters (Miyazaki \etal 2002) but also
for investigations into astrophysical effects during the non-linear
infall and growth on smaller scales. Full details of our calculations
are given in a series of papers by Rhodes \etal(2004), Massey \etal(2004b)
and Refregier \etal(2004b).

\begin{figure}
\centerline{\psfig{figure=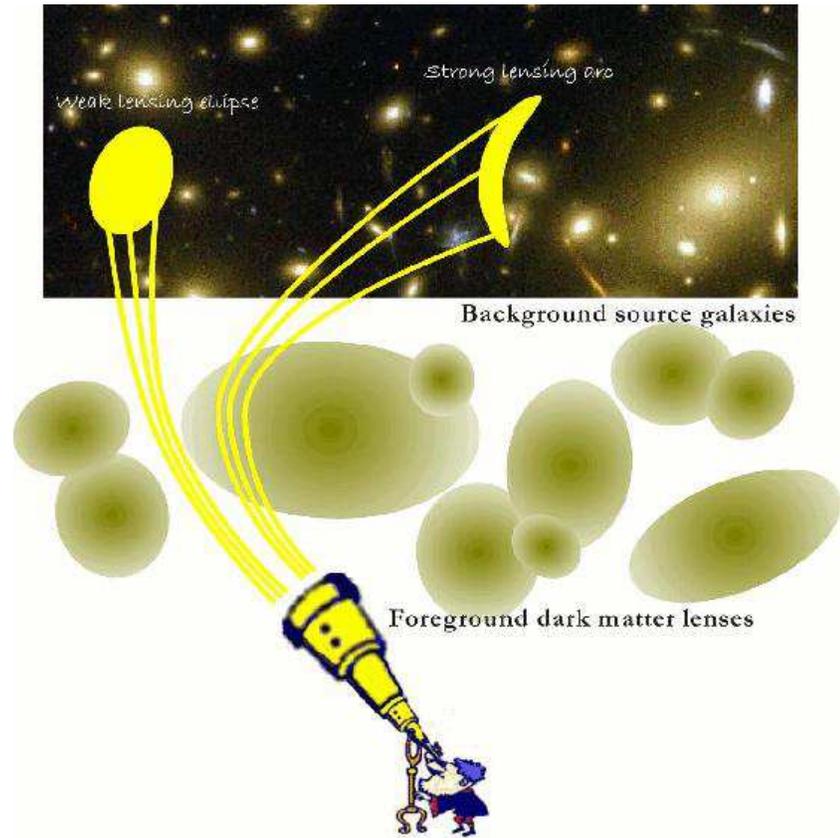,width=110mm}}
\caption{Illustration of the effect of gravitational lensing.
\hspace{2in}Foreground dark
matter haloes differentially shear the light paths from background galaxies,
distorting their shapes into an ellipse (when averaged over many galaxy shapes)
or arcs near the centres of clusters.
\label{fig:cartoon}}
\end{figure}

\section{Principles of Weak Lensing}

Gravitational lensing subjects the apparent images of background
galaxies to a distortion that is characterized by the distortion matrix
(see Figure 1 for an illustration, and Bartelmann \& Schneider
1999 for a detailed review)

\begin{equation}
 \label{eq:psi_def_theory}
 \Psi_{ij} \equiv \frac{\partial (\delta\theta_{i})}{\partial \theta_{j}}
           \equiv \left( \begin{array}{cc}
                                      \kappa + \gamma_{1} & \gamma_{2}\\
                                      \gamma_{2} & \kappa - \gamma_{1}\\
                         \end{array} \right) ~,
\end{equation}

\noindent where $\delta \theta_{i}({\mathbf \theta})$ is the deflection vector
produced by lensing on the sky. The convergence $\kappa$ describes overall
dilations and contractions, and is proportional to the projected mass along the
line of sight. The shear $\gamma_{1}$ ($\gamma_{2}$) describes stretches and
compressions along (at $45^{\circ}$ from) the $x$-axis.

The distortion matrix is directly related to the matter density fluctuations
along the line of sight by

\begin{equation}
\label{eq:lensing}
\Psi_{ij} = \int_{0}^{\chi_{h}} d\chi ~g(\chi)
            \partial_{i}\partial_{j} \Phi~,
\end{equation}

\noindent where $\Phi$ is the Newtonian potential, $\chi$ is the
comoving distance, $\chi_{h}$ is the comoving distance to the horizon,
and $\partial_{i}$ is the comoving derivative perpendicular to the
line of sight. The radial weight function $g(\chi)$ reflects the
fact that a lens is most effective when placed approximately
half-way between the source and the observer. It is given by

\begin{equation}
g(\chi) = 2 \int_{\chi}^{\chi_{h}} d\chi'~n(\chi')
   \frac{r(\chi)r(\chi'-\chi)}{r(\chi')} ~,
\end{equation}

\noindent where $r$ is the comoving angular-diameter distance. The function
$n(\chi)$ is the distribution of the galaxies as a function of the comoving
distance $\chi$ from the observer and is assumed to be normalized as $\int
d\chi n(\chi)=1$.

Galaxies have an intrinsic distribution of shapes, but distant ones are
randomly oriented in the absence of lensing. When a shear is applied to the
galaxies in a particular patch of the sky, they become coherently distorted and
their average shape changes from a circle to an ellipse. This observed
ellipticity can be converted into the applied shear using a method like KSB,
which properly takes into account their size distribution and radial profiles.
Thus the cosmic shear field is an observable. By inverting the lensing equation
(eqn.~2), the shear map can be converted into a map of the projected mass
$\kappa$ and, therefore, of the dark matter distribution. 

\section{Cosmological models with Dark Energy}

We consider a cosmological model with a matter component and a dark
energy (or ``quintessence'') component with present density parameters
$\Omega_{m}$ and $\Omega_{q}$, respectively.  The equation of state of
the dark energy is parametrized by $w=p_{q}/\rho_{q}$, which we assume
to be constant and is equal to $-1$ in the case of a cosmological
constant. The evolution of the expansion parameter $a$ is given by the
Hubble constant $H$ through the Friedmann equation

\begin{equation}
H=\frac{\dot{a}}{a}=H_{0}\left( \Omega_{m} a^{-3} + \Omega_{q}
  a^{-3(1+w)} + \Omega_{\kappa} a^{-2} \right)^{\frac{1}{2}} ~,
\end{equation}

\noindent where $\dot{a}=da/dt$ and the total and curvature density parameters
are $\Omega$ and $\Omega_{\kappa}=1-\Omega$, respectively. The present value of
the Hubble constant is parametrized as $H_{0}=100 h$ km s$^{-1}$ Mpc$^{-1}$.

As a reference model, we consider a fiducial $\Lambda$CDM cosmology with
parameters $\Omega_{m}=0.30$, $\Omega_{b}=0.047$, $n=1$, $h=0.7$, $w=-1$,
consistent with the recent CMB measurments from the WMAP experiment (Spergel
\etal2003). In agreement with this experiment, we assume that the universe is
flat, {\it i.e.} that $\Omega=\Omega_{m}+\Omega_{q}=1$. 

Dark energy has several effects on weak lensing statistics (Ma \etal1999).
First, it modifies the expansion history of the universe $a(t)$. As a result,
both the angular-diameter distance and the growth rate of structures are
modified. The latter effect is amplified by the non-linear evolution of
structures. In some quintessence models, dark energy also modifies the linear
power spectrum on large scales. We will ignore that effect since these scales
are not easily probed by weak lensing surveys.

\section{Cosmic shear statistics}
The properties of the cosmic shear field can be quantified by several
statistics. The most basic statistic is the weak lensing power
spectrum. This is the equivalent in Fourier space of the shear-shear
correlation functions, and is given by ({\it e.g.}~Bartelmann \&
Schneider 1999; Hu \& Tegmark 1999; see Bacon \etal2001 for
conventions)

\begin{equation}
C_{\ell} = \frac{9}{16} \left( \frac{H_{0}}{c}
\right)^{4} \Omega_{m}^{2}
  \int_{0}^{\chi_h} d\chi~\left[ \frac{g(\chi)}{a r(\chi)} \right]^{2}
  P\left(\frac{\ell}{r}, \chi\right) ~,
\label{eq:cl}
\end{equation}

\noindent where $r(\chi)$ is the comoving angular diameter distance, and
$\chi_{h}$ corresponds to the comoving radius to the horizon. Considerable
uncertainties remain for the non-linear power spectrum $P(k,z)$ in quintessence
models (see discussion in Huterer 2001). Here, we will use the prescription of
Peacock \& Dodds (1996) to evaluate it from the linear power spectrum. The
growth factor and COBE normalization for arbitrary values of $w$ can be
computed using the fitting formul\ae\ from Ma \etal(1999). 

\begin{figure}
 \centerline{\psfig{figure=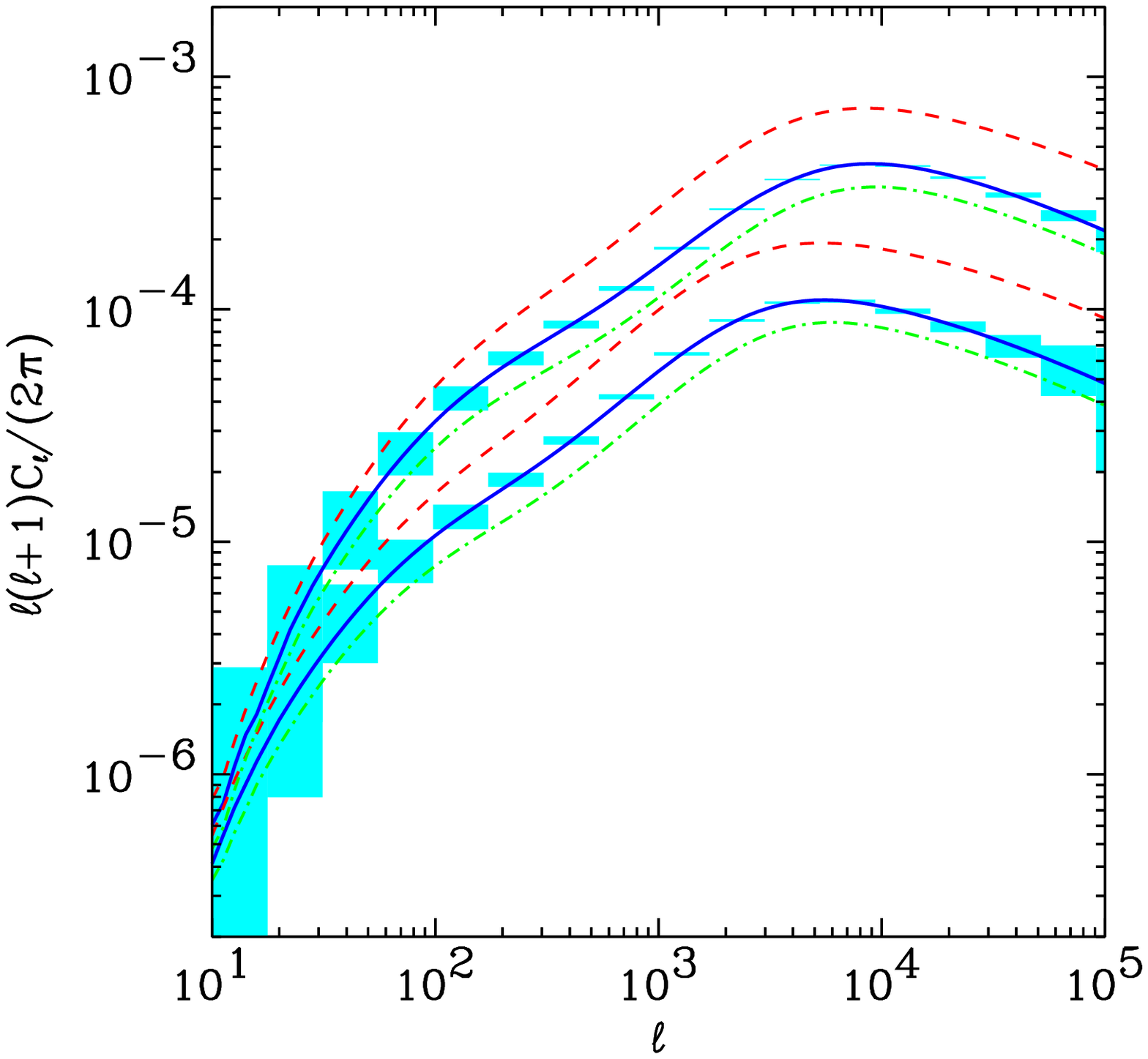,width=65mm}~~
 \psfig{figure=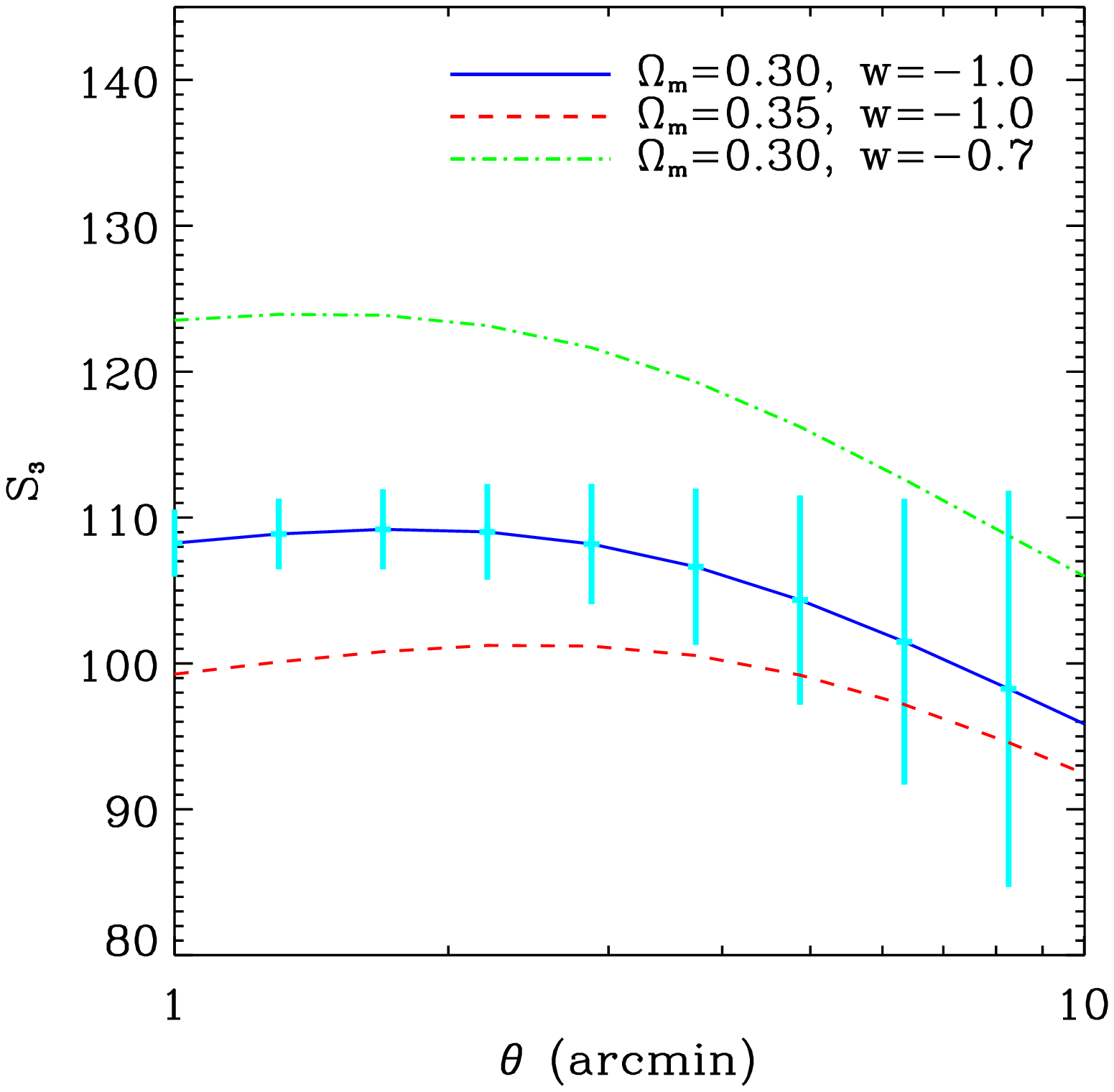,width=65mm}}
 \caption{Weak lensing statistics for different cosmological
 models. Variations of 17\% in $\Omega_m$ and of 30\% in $w$ about
 the fiducial $\Lambda$CDM model are displayed. Left: the weak lensing
 power spectrum $C_{l}$ for two galaxy samples with median redshifts of
 0.96 and 1.73. Right: skewness $S_{3}$ of the convergence $\kappa$ as a
 function of scale $\theta$. In each case, the expected band-averaged
 $1\sigma$ error bars for the SNAP weak lensing surveys are shown as blocks.
 (Adapted from Refregier \etal 2004b).} \label{fig:statistics}
\end{figure}

The left-hand panel of figure 2 shows lensing power spectra for the fiducial
$\Lambda$CDM model with background galaxies in two different shells, with
median redshifts of 0.96 and 1.73. Deviations from the model are also shown,
corresponding to variations in $\Omega_{m}$ and $w$. All models shown are COBE
normalised.  Note that non-linear corrections are dominant for $\ell \ga 100$.

Non-linear gravitational instability is known to produce non-Gaussian
features in the cosmic shear field. The power spectrum therefore does
not contain all the information available from weak lensing. We
consider the most common measure of non-Gaussianity, namely the
skewness $S_{3}$ which is defined as ({\it e.g.}~Bernardeau \etal1997)

\begin{equation}
 S_{3}(\theta) \equiv \frac{\langle \kappa^{3}
 \rangle}{\langle \kappa^2 \rangle^2} ~,
 \label{eq:s3}
\end{equation}

\noindent where $\kappa$ is the convergence which can be derived from the shear
field $\gamma_{i}$ and the angular brackets denote averages over circular
top-hat cells of radius $\theta$. The denominator is the square of the
convergence variance which is given by

\begin{equation} 
 \langle \kappa^{2} \rangle = 
 \langle \gamma^{2} \rangle \simeq 
 \frac{1}{2\pi} \int d\ell~\ell C_{\ell}|W_{\ell}|^{2} ~,
\end{equation} 

\noindent where $W_{\ell} \equiv 2J_{1}(\ell\theta)/(\ell\theta)$ is the window
function for such cells and $C_{\ell}$ is the lensing power spectrum given by
equation~(5). To evaluate the numerator $\langle \kappa^{3}
\rangle$ of equation~(6), we use the approximation of Hui (1999)
who used the Hyperextended perturbation theory of Scoccimarro \& Frieman
(1999). While more accurate approximations for third order statistics now exist
(see van Waerbeke \etal2001 and references therein), the present one suffices
for our present purpose.

The right-hand panel of figure 2 shows the skewness as a function of scale for
the same cosmological models considered in the left-hand panel. The skewness is
only weakly dependent on the angular scale $\theta$, but depends more strongly
on $\Omega_{m}$ and $w$.

\section{SNAP: a wide field imager from space} \label{snap}

The {\it Supernova/Acceleration Probe} (SNAP) satellite has a planned
launch date in 2010. The latest design of this wide-field 2m space telescope is
shown in figure 3. Many of the stringent optical requirements for following
the light curves of faint supernov\ae~are compatible with the desired
instrumental properties for measurements of weak lensing. Indeed, most of
SNAP's limitations and trade-offs will be born by any similar wide-field imager
from space. The detailed engineering models which are available for SNAP
therefore act as a useful baseline for a generic space mission which will
inevitably face similar engineering difficulties and reach similar solutions.

\begin{figure}
\centerline{\psfig{figure=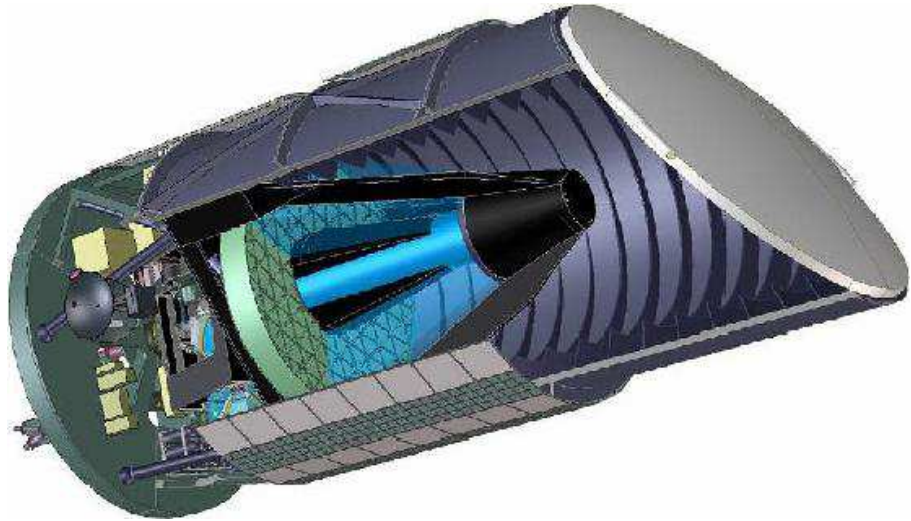,width=120mm}}
\caption{Cutaway view of the proposed SNAP satellite design,
showing the internal light baffles, the 2m primary mirror and the
three sturdy support struts of the secondary mirror. The solar
panels are fixed to the outside on one side of the craft, with a
radiator in the opposite direction. Their orientation with respect
to the sun is permanently maintained in order to minimise thermal
expansions and contractions that would otherwise induce optical
distortions, mimicking lensing shear. Image reproduced courtesy of
the SNAP collaboration.} \label{fig:cutaway}
\end{figure}

SNAP's 0.7 square degree field of view will be covered by a mosaic of nine
fixed filters spanning the optical and near IR from 350nm--1.7$\mu$m
(Perlmutter \etal~2003). These are arranged in such a pattern that scanning the
telescope horizontally or vertically across the sky will build up an image of a
contiguous survey region in all nine bands. At 800nm, the FWHM of the PSF will
be approximately 0.15\arcsec. The PSF, and any internal optical distortion
(that could mimic cosmic shear), will remain stable because of the satellite's
high and therefore thermally stable orbit. Indeed, the telescope will rarely
enter the shadow of the Earth and always maintain the same face pointing
towards the sun (Rhodes \etal 2004).

The SNAP survey strategy is divided into two primary missions. A deep
survey to ABmag 30.2 in R (for a point source at 5$\sigma$), will
cover 15 square degrees in all nine bands. This will be built up over
a period of 32 months, by stacking observations taken once every four
days in a search for type Ia supernov\ae~light curves. A second, wide
survey to ABmag 27.7 will cover 300 square degrees. This will be
gathered during a period of 5 months.

\begin{figure}
\centerline{\psfig{figure=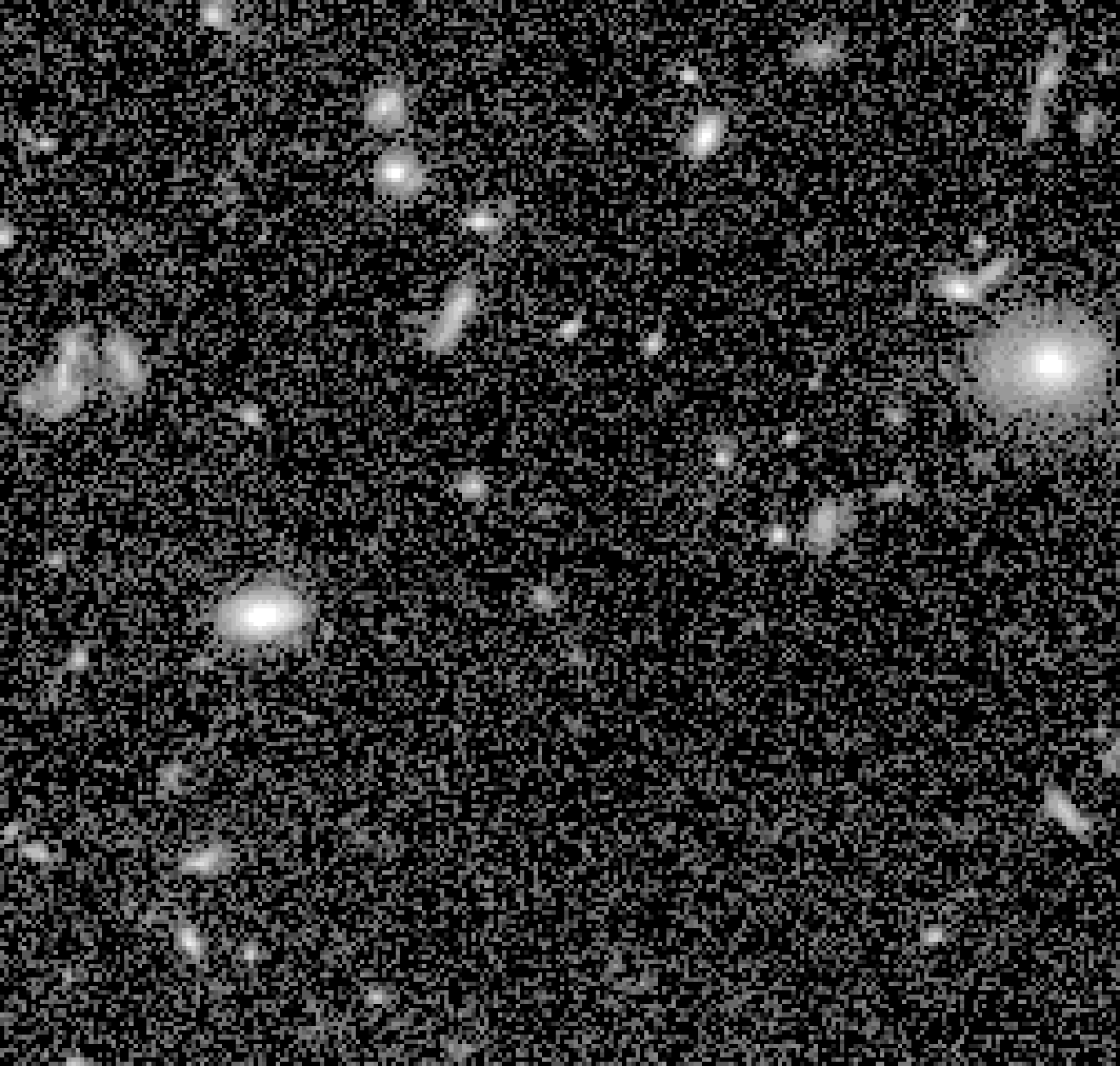,width=65mm}~~
\psfig{figure=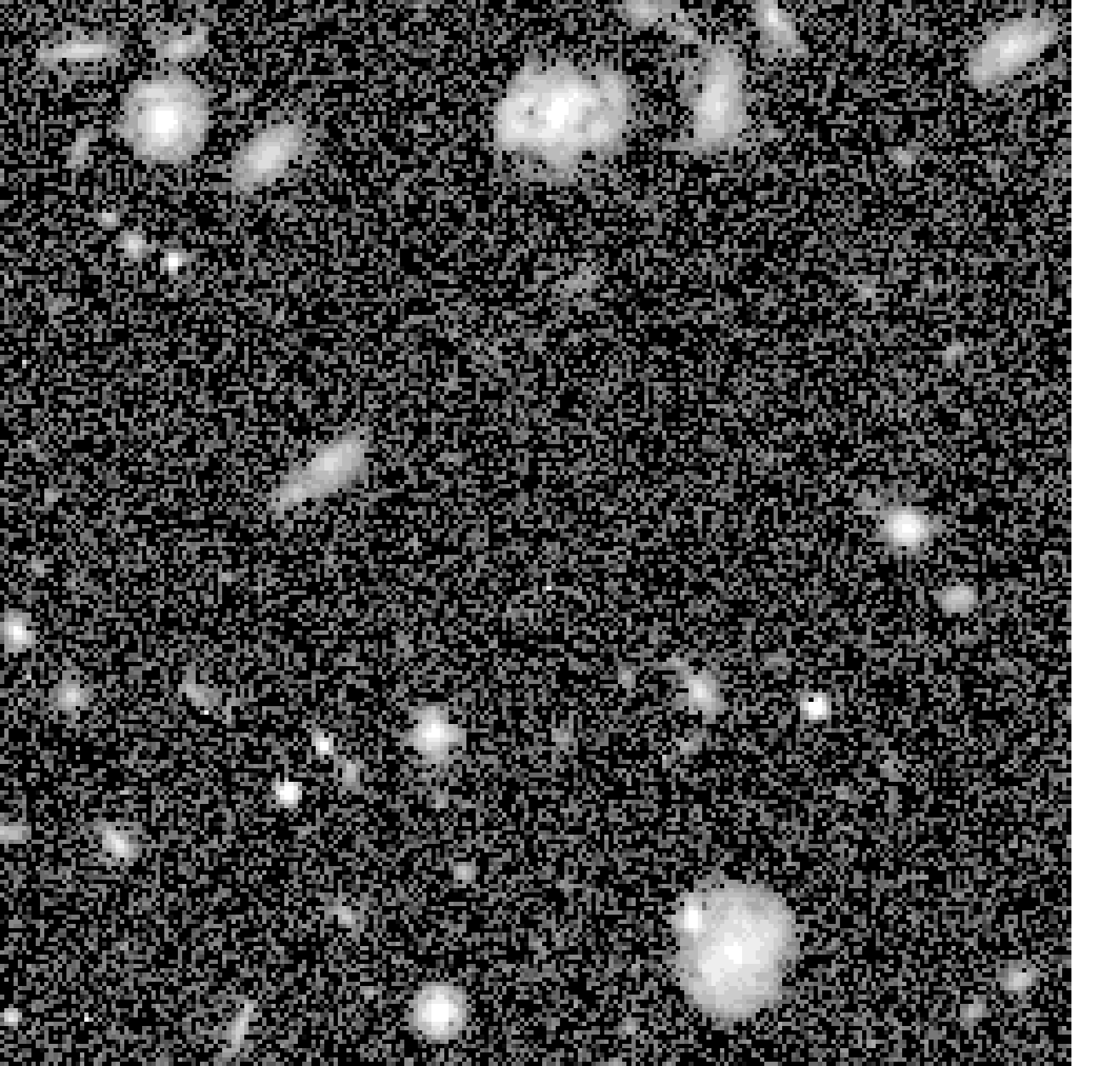,width=65mm}} \caption{An example of the
realistic simulated images created using the shapelet method of Massey
\etal (2004a). The left panel shows a $30\arcmin\times30\arcmin$
section of the real HDF. The right panel shows a section of a
simulated image with the same size and scale. The galaxies it contains
are no longer simple parametric models, but instead possess complex
morphologies that are statistically similar to those in observed
data. (Adapted from Massey \etal 2004a).} \label{fig:shims}
\end{figure}

Two major weak lensing applications are enabled by the SNAP
mission. The high number density of resolved galaxies (\simgt250
arcmin$^{-2}$) in the deep survey will be uniquely useful to construct
high-resolution mass maps and search for galaxy clusters by
mass. Complementing this, the wide survey is essential to reduce the
impact of cosmic variance on cosmological parameter estimation. For
this purpose, it is important that relatively high redshifts can still
be reached in even a modest exposure time from space -- and with
sufficient colours, including near IR, to provide accurate photometric
redshifts.

\section{Shapelet-based image simulations}
\label{shims}

The image simulation method of Massey \etal (2004a) was
developed to mimic deep, high resolution space-based data. The
simulations use the ``shapelets'' formalism of Refregier (2003a) to
create realistic galaxies of all morphological types and with all the
substructure and irregularity of faint HDF objects. The simulated
objects possess a known size, magnitude and input shear, and may be
used to calibrate the sensitivity of an instrument to weak
lensing.
 
\begin{figure}
\centerline{\psfig{figure=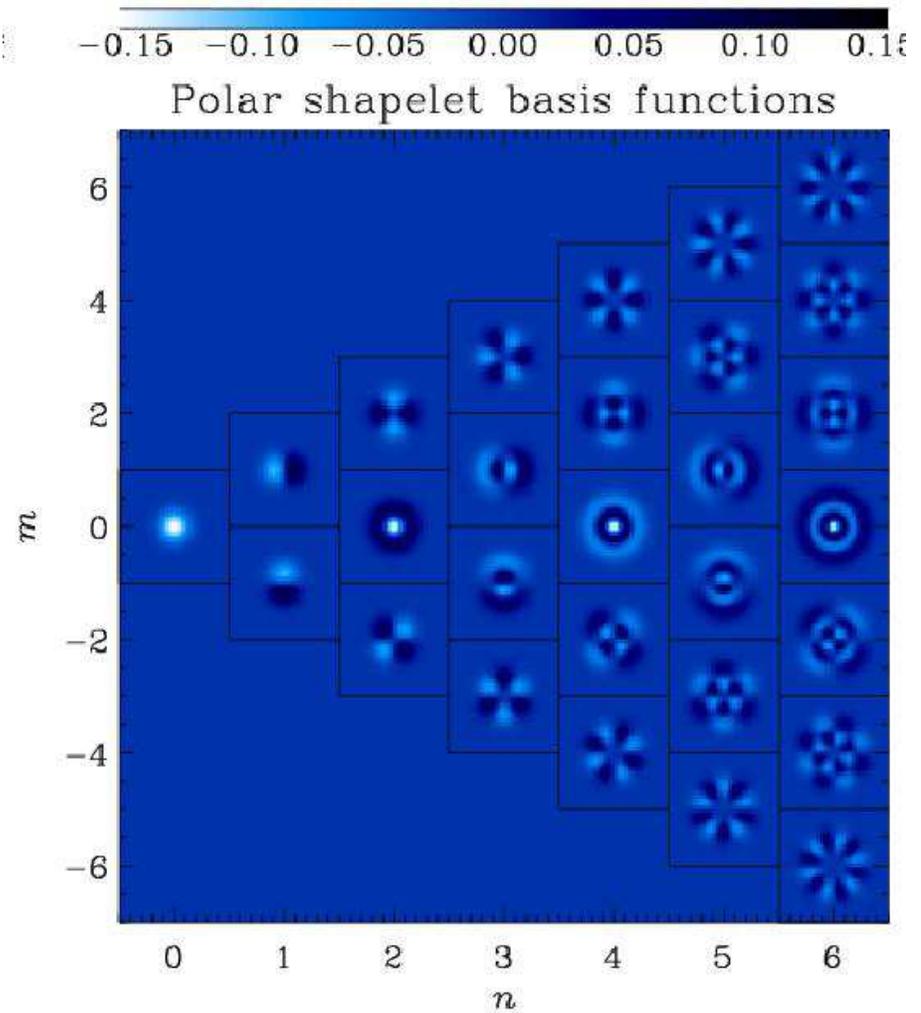,width=120mm}}
\caption{Polar
shapelet basis functions. These are a complete, orthonormal basis
set with which galaxy shapes can be modelled and regenerated later. 
Shapelets are mathematically convenient for many aspects of image analysis
and lie at the core of the image simulation method.}
\label{fig:sspace}
\end{figure}

As a specific case, we have manufactured simulated images to the design
specifications of the SNAP satellite. Since an object's response to shear is a
function of its overall shape, the presence of realistic and irregular galaxy
morphologies is an important advance over earlier work (Bacon \etal 2001, Erben
\etal 2001), which had used only azimuthally symmetric simulated galaxies with
radial profiles parametrized as de Vaucouleurs $r^{1/4}$ laws or exponential
discs. $30\arcmin\times30\arcmin$ sections of an example simulated image and a
similarly scaled section of the HDF are shown in figure 4.

\begin{figure}
\centerline{\psfig{figure=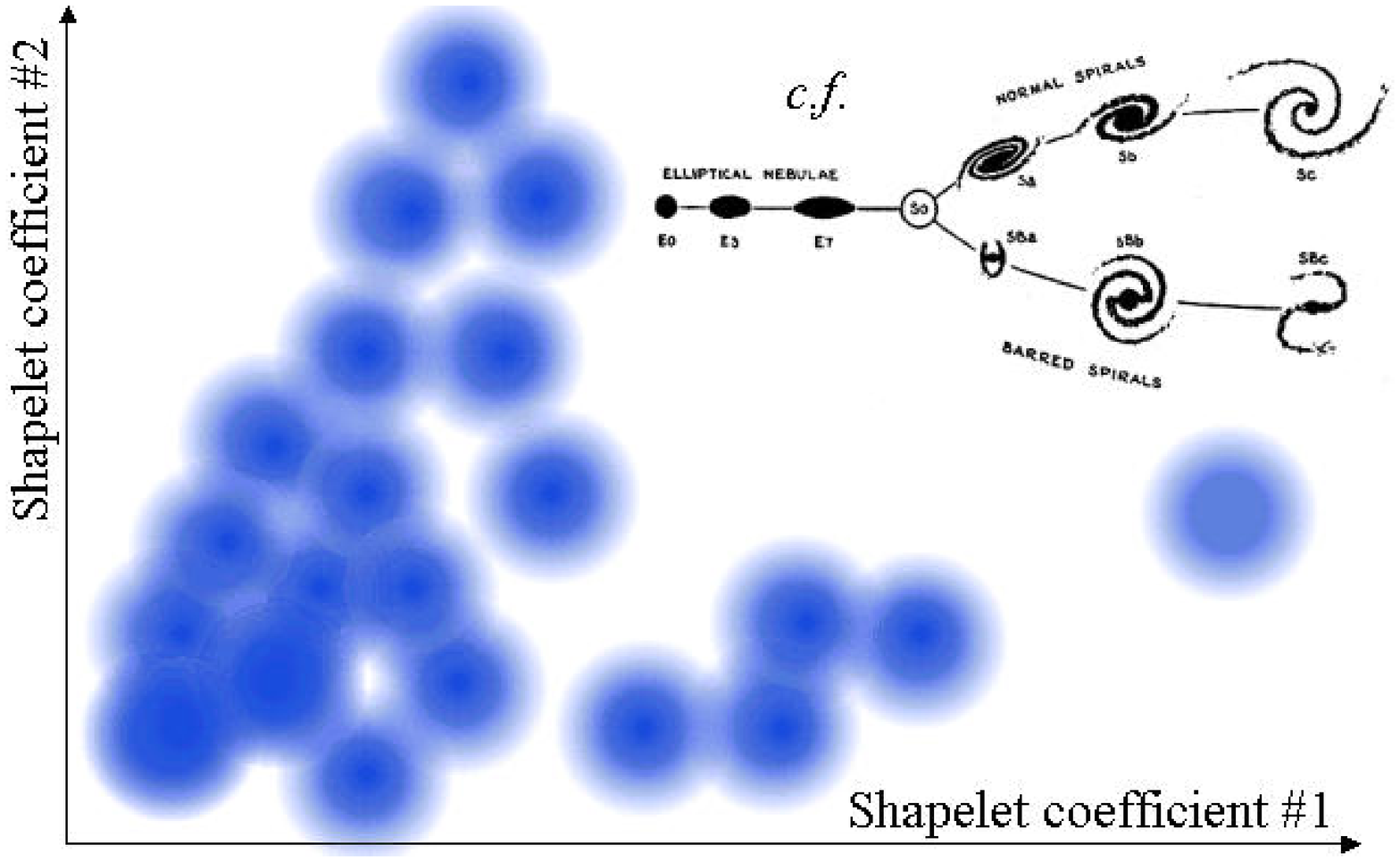,width=100mm}}
\caption{Idealised representation of a slice through shapelet
morphology parameter space, where each dimension corresponds to a
shapelet coefficient. The points represent the position of an observed
HDF galaxy in shapelet space. Their distribution encode the underlying
morphology distribution of real galaxies, like a multi-dimensional
Hubble sequence. This morphology distribution can be smoothed and
resampled to generate an unlimited number of unique, yet realistic
galaxy shapes for the image simulations.} \label{fig:sspace}
\end{figure}

To generate the image, the galaxies in the HDF-North and HDF-South are first
decomposed into shapelets. Shapelets are a complete basis set of
two-dimensional, orthonormal basis functions, constructed from Laguerre
functions multiplied by a Gaussian, as shown in figure 5. This basis is
mathematically convenient for many aspects of image manipulation and analysis,
including weak lensing shear and magnification. Linear combinations of shapelet
basis states, weighted by ``shapelet coefficients'', can be used to model any
localised shape, rather like Fourier or wavelet synthesis. The shapelet
coefficients are Gaussian-weighted multipole moments of the object, which can
be used for quantitative shape measurement.

\begin{figure}\label{fig:flowchart}
\centerline{\epsfig{figure=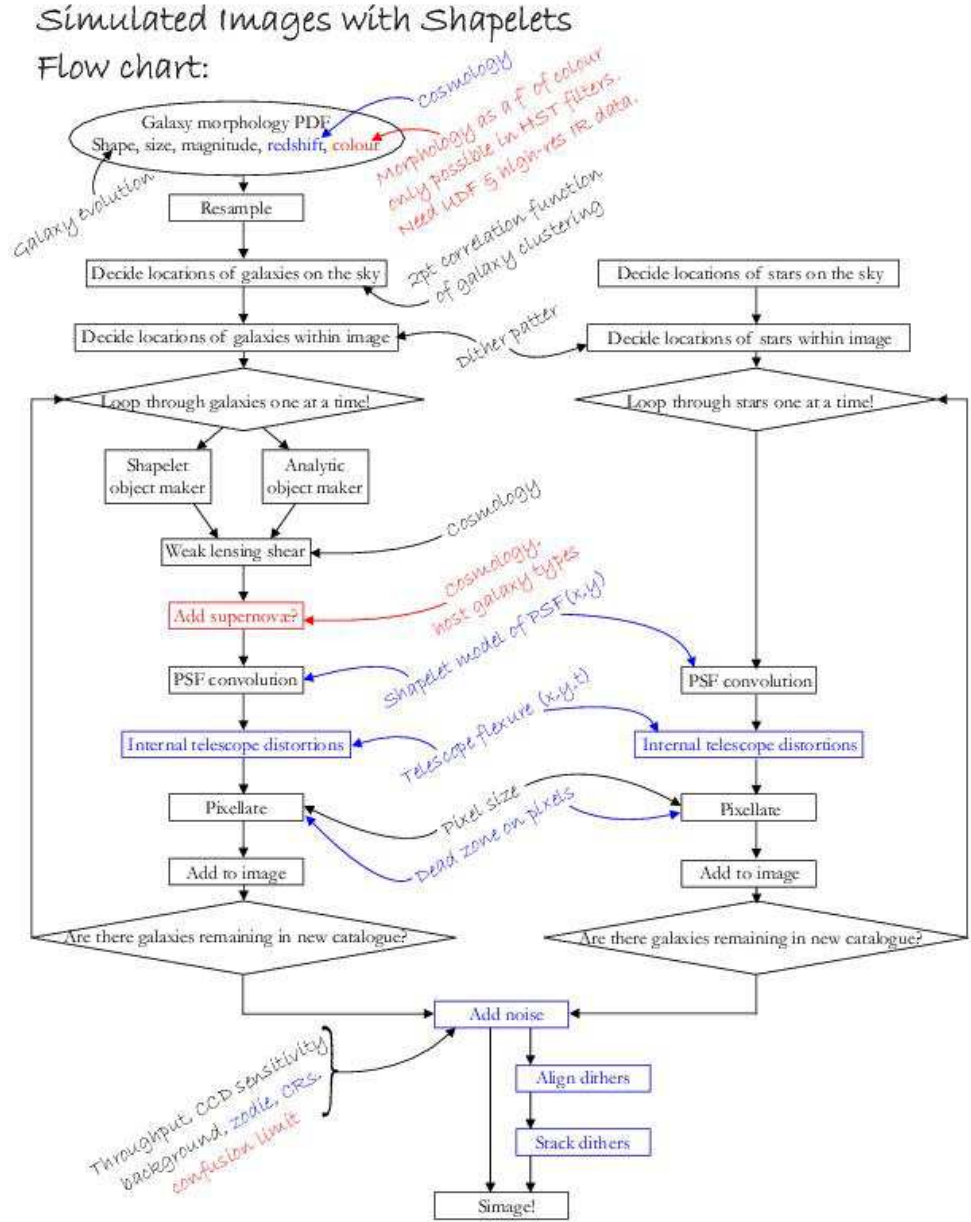,width=130mm}}
\caption{Flowchart showing the steps taken to produce a simulated
image. These steps mimic the processes acting on photons
en route from a distant galaxy to a telescope. The required
inputs from cosmology, an engineering model of the telescope and a
survey strategy are shown in a script font. Type Ia supernov\ae~could
also be added, to simulate the other aspect of the SNAP mission.}
\end{figure}

The shapelet coefficients of the real HDF galaxies populate an
$n$-dimensional vector space, where $n$ is the maximum number of
coefficients used. The vector space is illustrated in figure 6. It
is analagous to the Hubble tuning fork: some regions of shapelet
space represent spiral galaxies; other regions represent early-type
elliptical galaxies. The underlying distribution of galaxy
morphologies is only finitely sampled by the HDF, but may be recovered
by smoothing the distribution in shapelet space. Massey \etal (2004a)
justify this process by comparing SExtractor and
concentration/asymmetry morphology estimators of real data to the
consistent statistics of shapelet-generated galaxies in the
final simulated images.

An image of distant galaxies can then be manufactured by repeatedly
resampling the recovered morphology distribution. Figure 7
illustrates the steps taken to produce a realistic simulated image.
These steps mimic the processes acting on photons en route from
a distant galaxy to a telescope. First, the galaxies are sheared by
the input gravitational lensing signal. Then they are convolved with
the SNAP PSF and slightly distorted through a model of the telescope's
internal optics.  Typical observational noise is added at the
end. These images can then be fed into a data reduction and shear
measurement pipeline. As we show in the next section, comparing the
known input shear with the output of this pipeline helps both to
determine SNAP's sensitivity to weak lensing and to calibrate and
improve shear measurement algorithms.

\section{Weak Lensing Measurement}
\label{shims}

We have used the RRG shear measurement method to attempt to recover
the known level of shear input to the simulated images. Figure 8
shows the sensitivity to shear as a function of exposure time.

\begin{figure}
\centerline{\psfig{figure=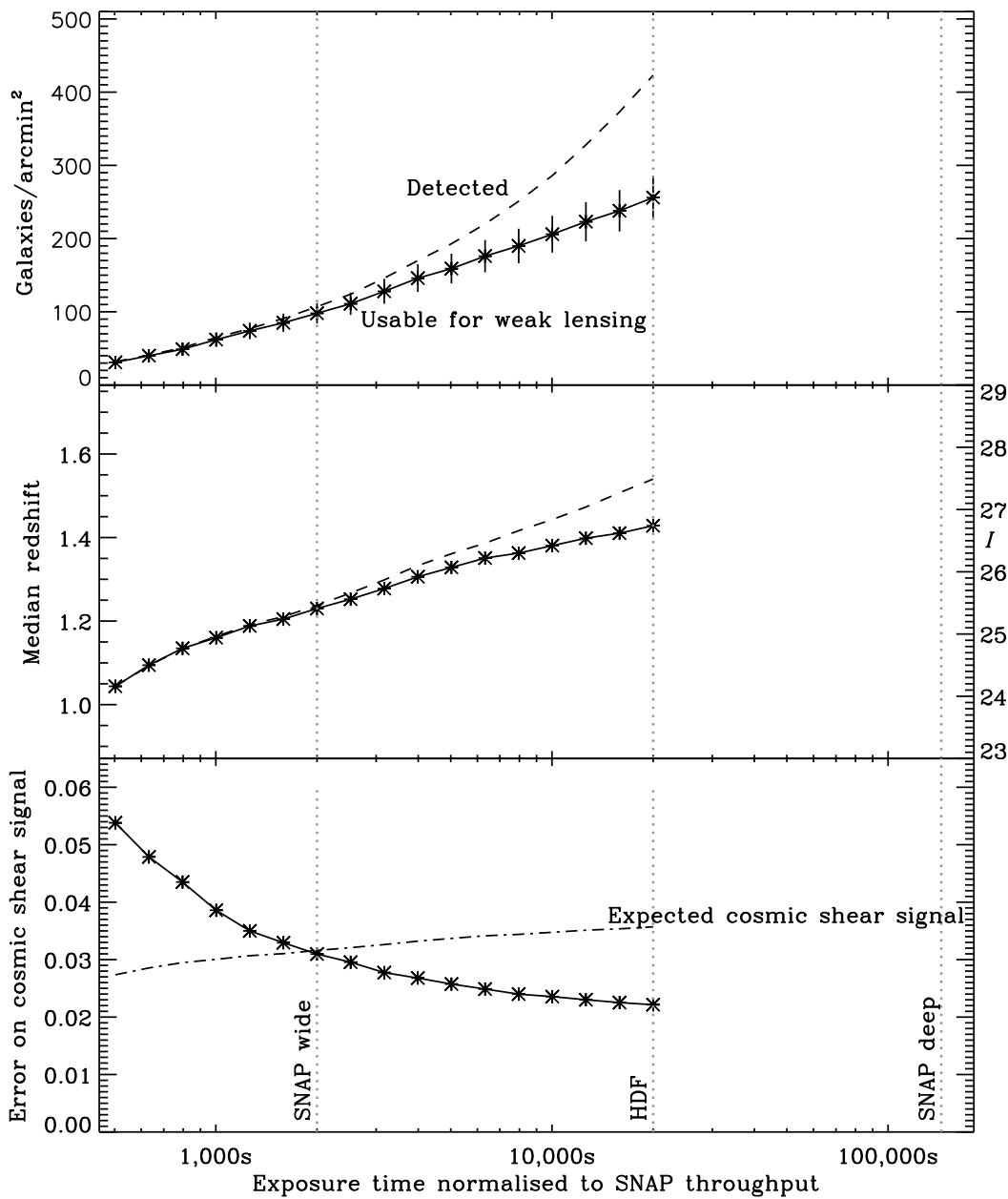,width=140mm}}
\caption{Sensitivity to weak lensing shear, as a function of exposure time on
SNAP.  Deep ACS images are awaited to model galaxies fainter than the current
limit of the HDF. Top panel: the number density of detected objects (dashed)
and of the subset sufficiently well resolved to be useful for shear measurement
(solid). Middle panel: the increase in redshift as a function of survey depth
for the two galaxy populations shown above. Bottom panel: the predicted error
on cosmic shear measurements in bins of 1~arcmin$^2$, using SNAP. Shown dotted
is the prediction of cosmic shear signal for a $\Lambda$CDM model. On 1'
scales, the SNAP wide survey will achieve an average S/N of unity: ideal for
making maps. (Adapted from Massey \etal 2004b).}
\label{fig:texp}
\end{figure}

There are three main advantages of using space-based images for weak lensing
measurements. Firstly, the smaller PSF means that there is a higher space
density of galaxies sufficiently resolved for their shears to be measured, and
shot noise is reduced on small scales. This is illustrated in the top panel of
figure 8. Although the number of {\it detected} galaxies (dotted line)
increases rapidly with increasing exposure time, the fraction of these which
are resolved (solid line) begins to drop as the size distribution of the faint
galaxies falls beneath the PSF size. Note that the simulations currently extend
only to the depth of the HDF. The slope of the number counts beyond this, and
the shape properties of these faint galaxies are unknown. We therefore await
deeper observations with the ACS to probe this regime.

The second advantage of space is that the galaxies resolved from space are
generally further away than those just resolved from the ground. The cosmic
shear signal is predicted to increase with survey depth due to the cumulative
lensing by more structures along the line of sight. The middle panel of figure
8 shows a prediction of the mean redshift of galaxies in the lensing sample
as a function of exposure time. This is calculated via a simple model of median
magnitude $vs$ median $z$, extrapolated from photometric observations of the
HDF (Lilly \etal 1996).

The third advantage is that the shape of each galaxy can be more reliably
measured. Not only is the S/N higher on each galaxy, with more resolved
morphological detail, but the correction for instrumental smearing and
distortion can be much more accurate because of SNAP's stable PSF. The PSF
stability of SNAP will be superior even to that of HST, which suffers from
telescope breathing due to the large changes in its temperature while passing
in and out of direct sunlight during each orbit.

The bottom panel of figure 8 shows the resulting sensitivity to lensing of a
space-based mission like SNAP. The error on the shear of galaxies binned into
1~arcmin$^2$ patches on the sky is shown as a function of exposure time. The
dotted line shows the expected cosmic shear signal, which increases with
exposure time due to the cumulative lensing by more and more objects as the
survey depth increases. The S/N on scales of 1~arcmin$^2$ is about 1 and
greater than about 1.8 for the wide and deep surveys respectively. As we
discuss in the next section, this offers great prospects for mapping the dark
matter.

In contrast to this precision from space, consider the limitations from a
ground-based telescope. The ESI camera on Keck has proven itself to have
minimal levels of internal optical distortion and telescope flexure, as
required for high precision weak lensing measurements. The size of the primary
mirror also lets it reach useful depths in exposures of only ten minutes (Bacon
\etal 2002). However, even in this short time, variations in the ground-based
PSF seriously affect weak lensing observations with any ground-based
telescope.  Figure 9 shows examples of these changing atmospheric patterns, all
taken with ESI on the same night. The line segments show the direction and
ellipticity of the PSF, as measured from stars. Bacon \etal (2001) and Erben
\etal (2001) demonstrated that the smearing of object shapes can be corrected
at a level of $\sim90$\% using the KSB method, and the corrected shapes of our
stars are also shown in figure 9.

From space, the raw PSF ellipticities before correction are
around the level on Keck after correction (Rhodes \etal
2004). Coupled with more stable optics, and newer correction methods,
this will result in a reduction in overall systematic contamination by
at least an order of magnitude.

\begin{figure} \centerline{\psfig{figure=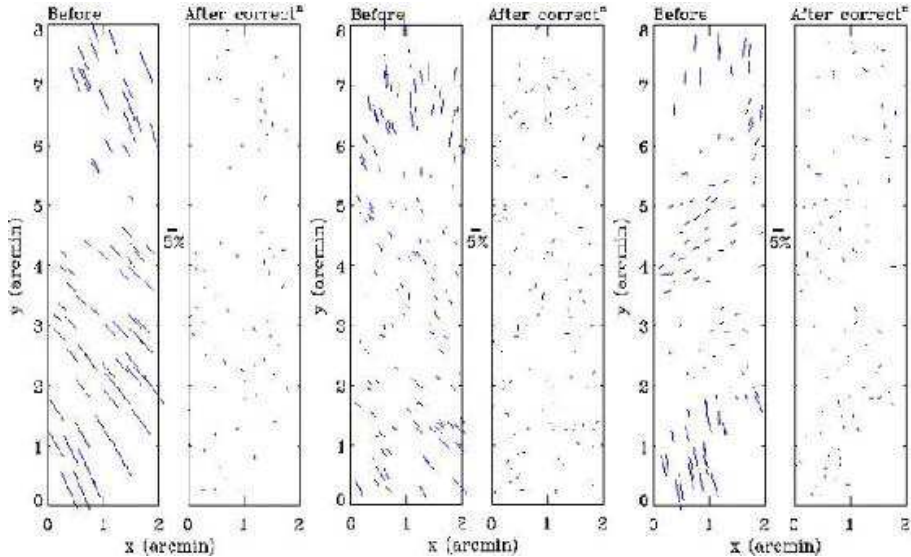,width=120mm}}
\caption{The constantly changing PSF anisotropy patterns from one night on
Keck. Although systemtics effects are well-controlled on Keck, and exposure
times can be short, the seeing effects of the atmosphere are unavoidable. 
Shown above are the amount and direction of PSF ellipticities as measured from
stars, before and after corrections. The patterns are not well understood but
are probably due to a combination of imperfect tracking, telescope flexure, and
atmospheric turbulence.  The galaxy shapes must be corrected for these effects
before shear measurements can be achieved. Plotted on the same scale, the
equivalent PSF ellipticities for SNAP would be barely visible.}
\label{fig:whisker} \end{figure}

\section{Cosmological constraints}

Two- and three-point statistics of the cosmic shear field in the wide SNAP
survey will be used to constrain cosmological parameters. Figure 2 shows the
effect of varying $\Omega_m$ and $w$ on the weak lensing power spectrum and
skewness. The lensing power spectrum is shown for 2 bins of galaxy redshifts
which can be derived from photometric redshifts. The error bars expected for
the SNAP wide survey (using $A=200$ deg$^{2}$, $\sigma_{\gamma}=0.31$ and
$n_{g}=100$ deg$^{-2}$) are displayed. The excellent precision afforded by SNAP
will easily distinguish the models shown, and thus detect small changes in the
properties of dark energy (Refregier \etal 2004b).

\begin{figure}
\centerline{\psfig{figure=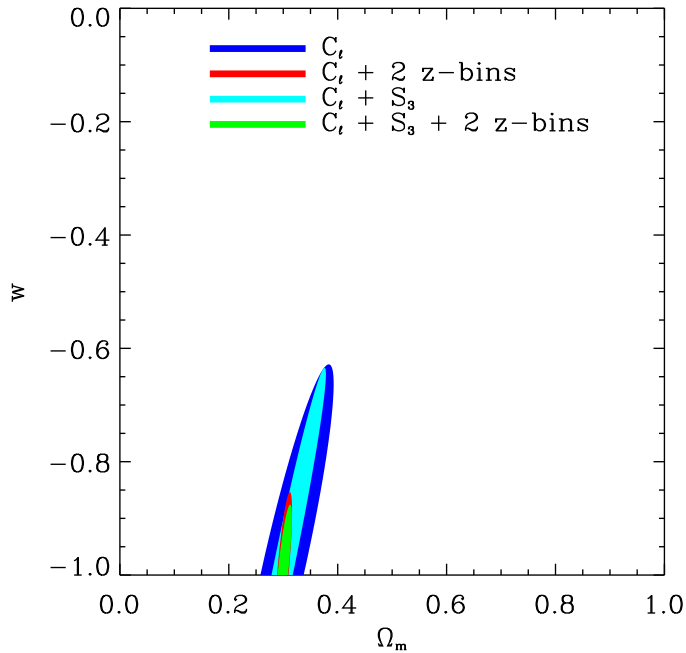,width=100mm}}
\caption{Predicted cosmological parameter constraints from SNAP weak lensing
(Refregier \etal 2004b). Contours show 68\% CL limits constraints derived from
the power spectrum $C_{l}$ using the SNAP wide survey ($A=300$ deg$^{2}$,
$\sigma_{\gamma}=0.31$ and $n_{g}=100$ deg$^{-2}$). From top to bottom, the
different colours show constraints with and without photometric redshifts, with
the skewness $S_{3}$ and with all of them combined. A COBE normalisation prior
was used. (From Refregier \etal 2004b).} 
\label{fig:coscons1}
\end{figure}

We can compute the constraints which can be set on cosmological
parameters using the Fisher matrix ({\it e.g.}~Hu \& Tegmark 1999)

\begin{equation}
 F_{ij}= - \left\langle \frac{\partial \ln {\mathcal L}}
           {\partial p_{i} \partial p_{j}} \right \rangle ~,
\end{equation}

\noindent where ${\mathcal L}$ is the Likelihood function, and $p_{i}$ is a set
of model parameters. The inverse ${\mathbf F}^{-1}$ provides a lower limit for
the covariance matrix of the parameters.

Figure 10 shows the constraints on the $\Omega_m$--$w$ plane that
will be possible from SNAP. The measurement of the weak lensing power
spectrum at two different redshifts (or ``redshift tomography'')
provides an important lever arm upon the growth of structure and the
evolution of the power spectrum, improving the constraints
signicantly. The addition of the skewness (at a single angular scale)
does not improve the constraints as much. Figure 11 compares the
constraints from weak lensing (using the power spectrum in two
redshift bins and the skewness) to that which can be derived from
current and the future SNAP supernova surveys (Perlmutter \etal1999,
2003). Note however that the supernov\ae~constraints shown include
systematic errors and a marginalisation over $w'$, the time derivative
of $w$. The SNAP weak lensing survey will therefore provide
constraints on the dark energy which are comparable with and
complementary to those from the SNAP supernova survey.

\begin{figure}\label{fig:coscons2}
\centerline{\psfig{figure=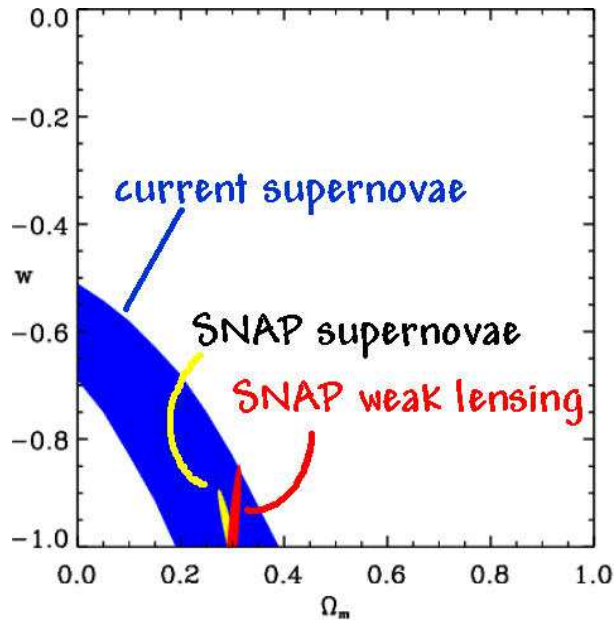,width=80mm}}
\caption{A comparison of the constraints derived from weak lensing with SNAP
(using the combined skewness and tomography statistics as before, but without
the COBE prior), from that with current and future SNAP supernova surveys
(Perlmutter \etal1999, 2003). (Adapted from Refregier \etal 2004b).} 
\end{figure}

\begin{figure}
 \centerline{\psfig{figure=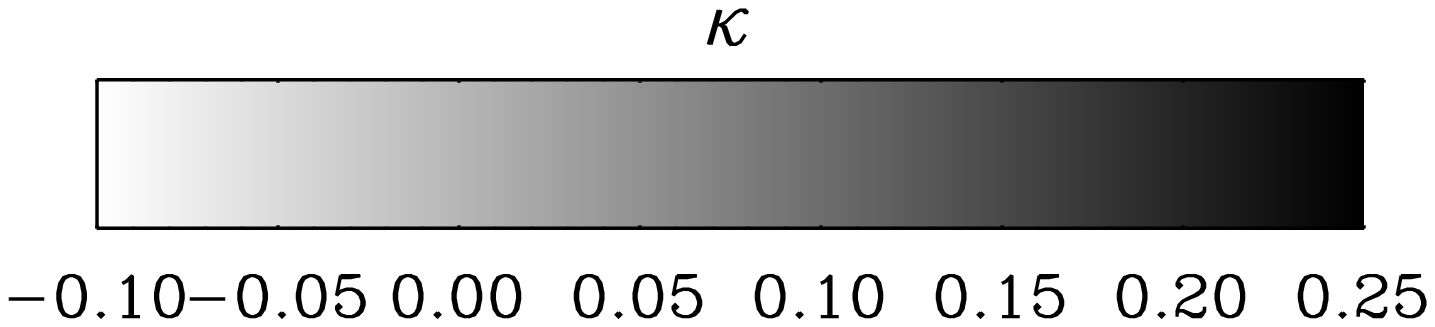,width=75mm}~~~~~} ~\\
 \vspace{-62mm}
 \centerline{\psfig{figure=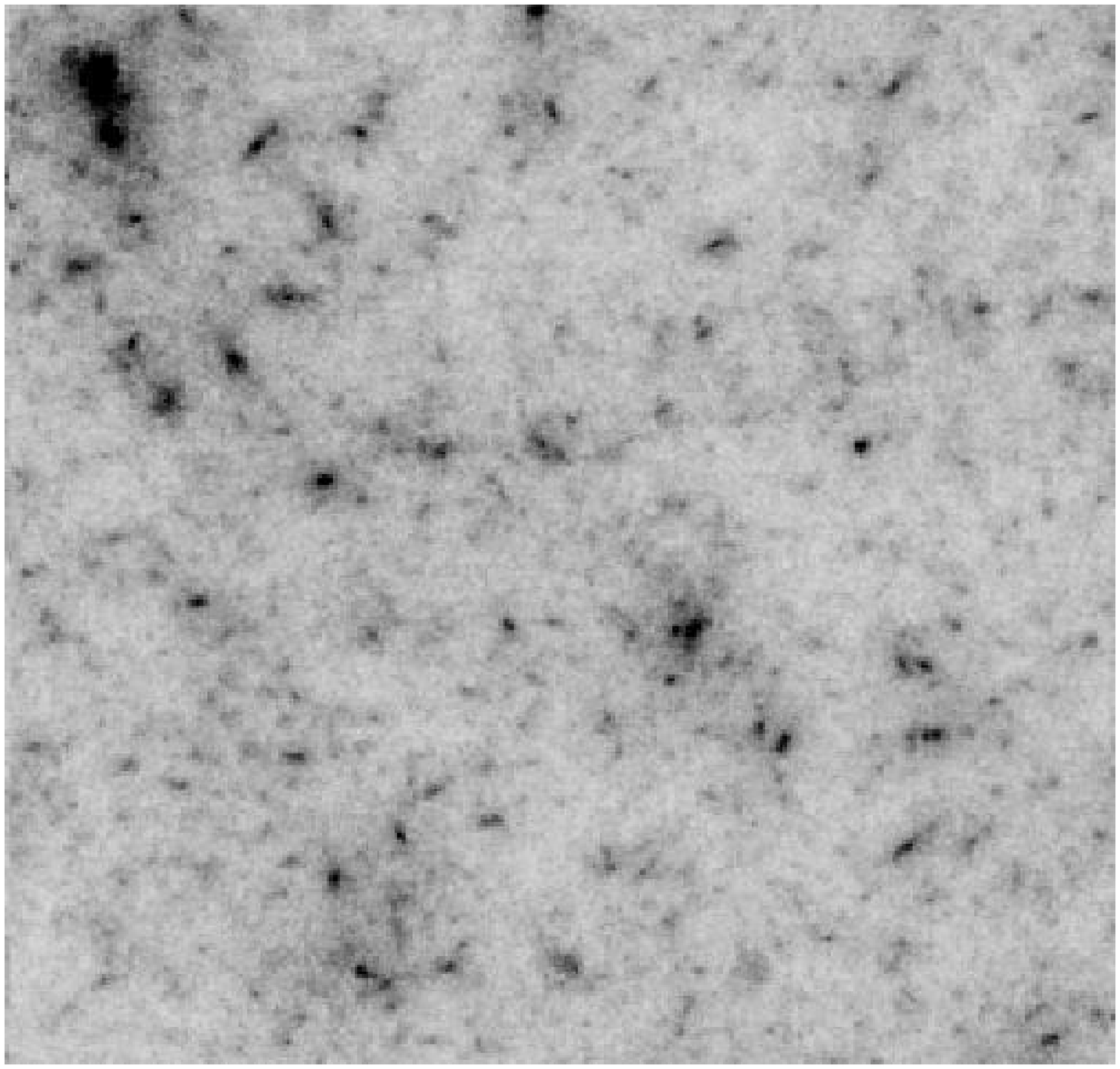,width=65mm}~~\psfig{figure=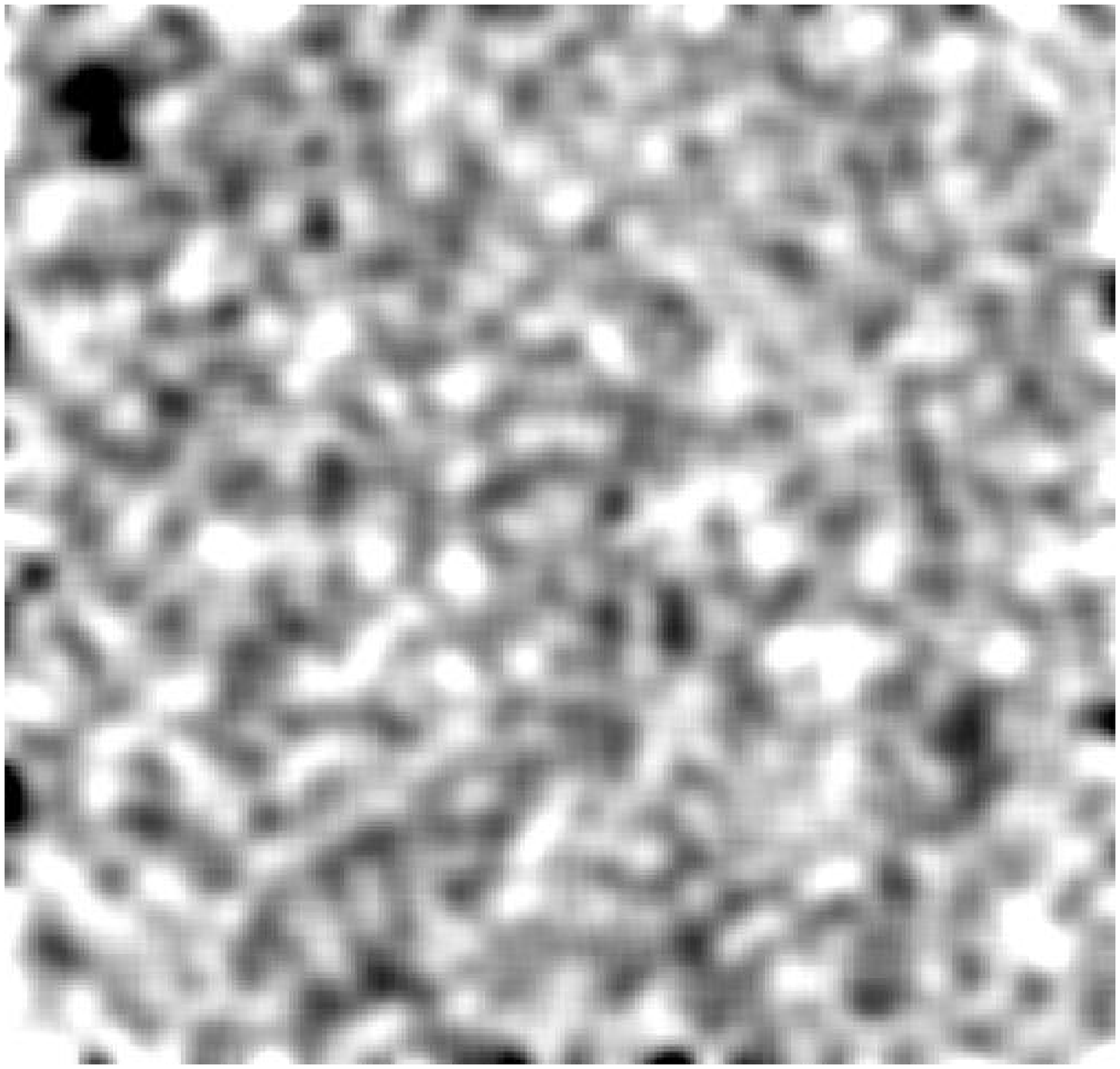,width=65mm}}~\\
 \vspace{-2mm}
 \centerline{\psfig{figure=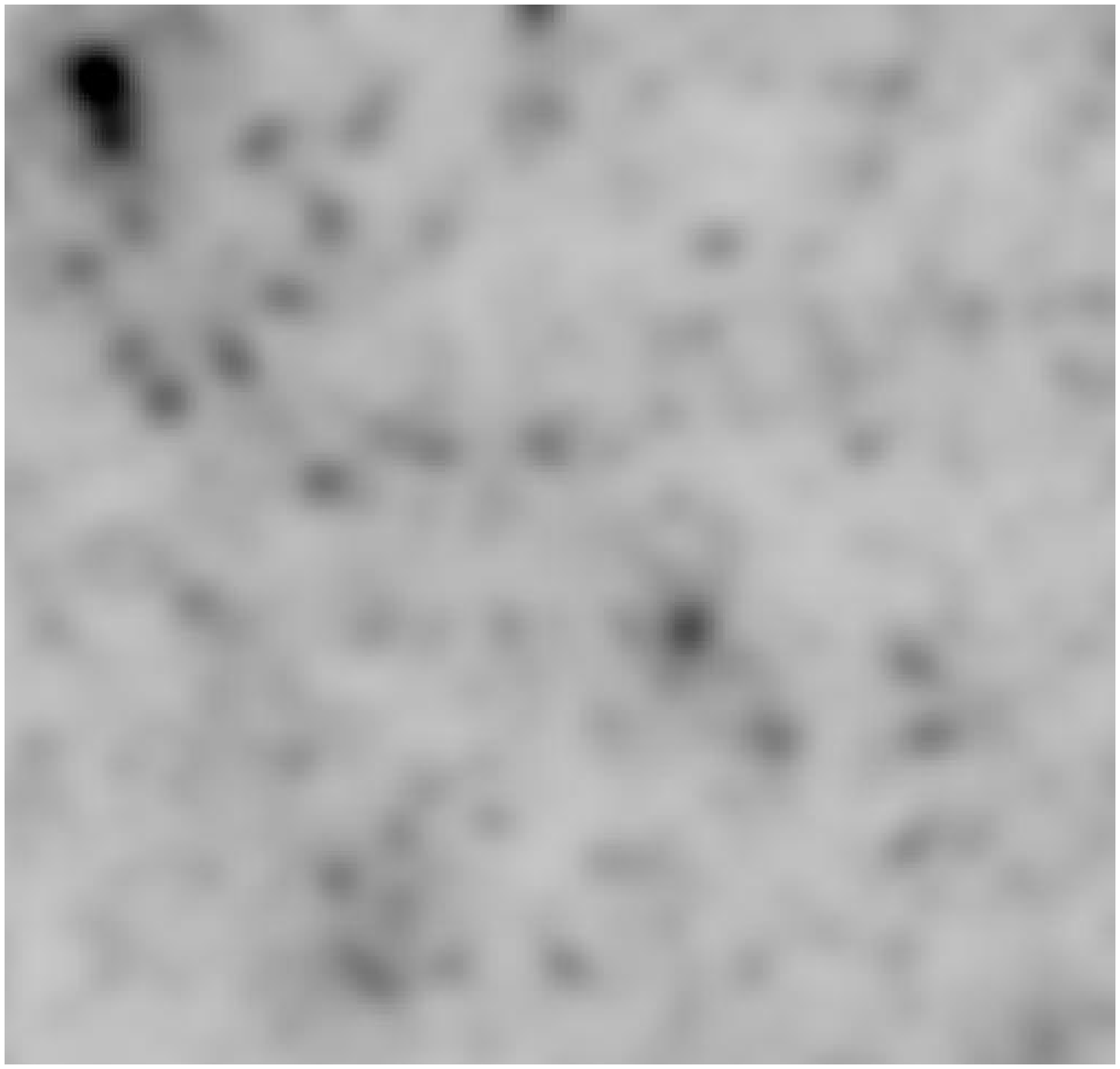,width=65mm}~~\psfig{figure=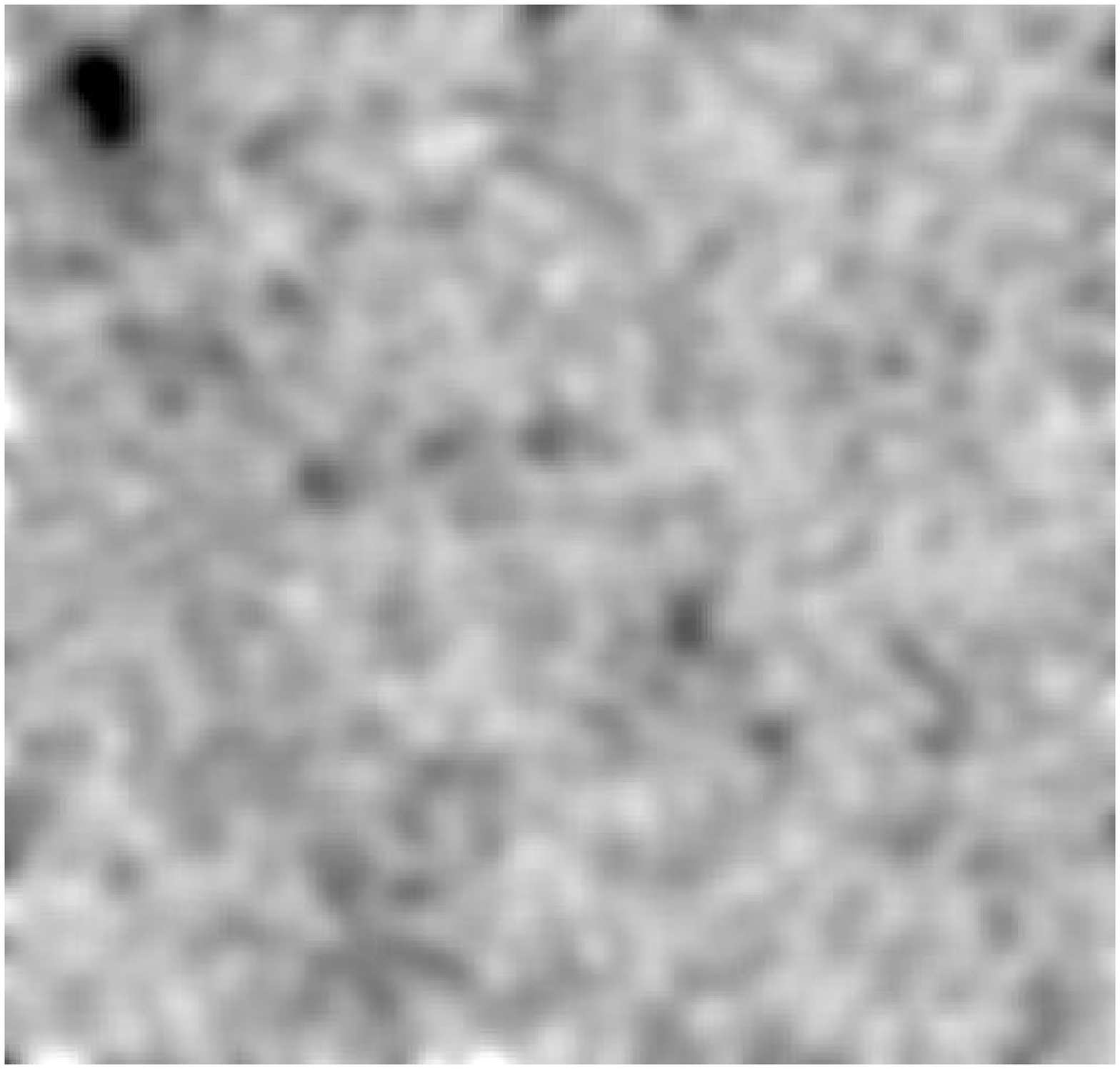,width=65mm}}~\\

 \caption{Reconstructed 2-dimensional maps of convergence $\kappa$, each
 $30\arcmin\times30\arcmin$. Convergence is proportional to the total matter
 density along the line of sight, and can be deduced from the shear field.
 Top-left: simulated (noise-free) convergence map derived by raytracing through
 an SCDM N-body simulations of large-scale structure from Jain, Seljak \& White
 (2000), in which the sources are assumed to lie at $z=1$. Bottom-left: same
 map but smoothed with a Gaussian kernel with a FWHM of 1 arcmin. Top-right:
 Reconstruction of the convergence map with a noise realization corresponding
 to ground-based observations. Statistics are taken from a WHT survey by Bacon
 {\it et al.}~2002 ($n_{g}=25$ arcmin$^{-2}$ and $\sigma_{\gamma}=0.39$).
 Bottom-right: convergence map with much-improved recovery from the expected
 noise properties of the SNAP wide survey ({\it i.e.}~$n_{g}=105$ arcmin$^{-2}$
 and $\sigma_{\gamma}=0.31$).}

 \label{fig:maprec}
\end{figure}

\section{Dark Matter Mapping}

From the observed shear field, one can reconstruct maps of the lensing
convergence $\kappa$, which is proportional to the total mass projected along a
given line of sight. The resolution of the maps depends on the size of the
spatial element within which shear or convergence can be accurately measured to
a S/N$\sim1$. The SNAP wide survey has been tailored to resolve the shapes of
$\sim100$ background galaxies per square arcminute, over 300 square degrees.
For the instrumental sensitivity to shear read from figure 8, this will achieve
maps with a resolution of 1 arcmin$^2$ pixels ($\sim$250kpc at $z=0.3$).
Extrapolating our simulations beyond the depth of the HDF, the SNAP deep survey
will exceed even this density requirement by a factor of three or four (Massey
\etal2004b). Thus, SNAP will open up a new regime of dark matter mapping,
allowing direct comparison to be made between mass and light on fine scales
over a very wide field of view.

In figure 12 we demonstrate the precision with which SNAP will be able to map
the dark matter in an SCDM simulation from Jain, Seljak \& White (2000).  In
this simulation, the source galaxies are assumed to lie on a single plane at
$z=1$. Whilst only the most massive overdensities can be distinguished in
ground-based shear data, the recovery is much improved from space. This is
mainly due to the three- to four-fold increase in number density of resolved
galaxies in even the SNAP wide survey. The statistical correlation of such maps
will allow the bias between mass and light to be examined to high accuracy over
a large range of scales.

Furthermore, simultaneously combining shear estimation with photometric
redshifts will permit the use of a recently formulated direct 3D lensing
inversion (Taylor 2001; Heavens in this volume). This method directly recovers
the full 3D mass distribution without the need to slice projected maps into
redshift bins. Applied to the SNAP deep survey, this technique will detect mass
overdensities with a $1\sigma$ sensitivity lower than $10^{13}M_{\sun}$
at $z=0.25$. An unbiased, mass-selected cluster catalogue will trace the growth
of mass structures in the universe (see Miyazaki \etal2002 and references
therein).

\section{Conclusions}

A space-based, wide-field imager is ideal for weak lensing
measurements. Indeed, a systematic floor due to atmospheric seeing
will be reached in the next generation of lensing surveys. This is
due both to the isotropic PSF smearing that reduces the number of
resolved galaxies for which shape measurements are
possible; and anisotropic smearing which cannot be perfectly
corrected when it varies from one exposure to the next. The
proposed SNAP satellite will have both the wide field needed to
survey large, representative cosmic volumes and the low level of
instrumental systematics affecting galaxy shapes.

Through a dual wide and deep survey strategy, weak lensing with
SNAP will be able to produce unique maps of the dark matter
distribution, on small scales and in both 2D and 3D. Weak lensing
with SNAP will be able to produce mass-selected cluster catalogues
to a $1\sigma$ detection threshold of $10^{13}M_{\sun}$ at
$z=0.25$.

Weak lensing primarily measures the distribution of the dark matter, but it
is also a powerful probe of dark energy. A change in the equation of state $w$
of dark energy modifies the growth rate of structures and the angular-diameter
distance. The resulting changes in the cosmic shear statistics will be easily
detectable with future weak lensing surveys. In particular,
weak lensing measurements with SNAP will be able to independently constrain
cosmological parameters $\Omega_M$, $\sigma_8$ and $w$ at a level comparable
with and somewhat orthogonal to those from future supernova searches and CMB
experiments.

\newpage
\acknowledgements The authors would like to thank David Valls-Gabaud
and Jean-Paul Kneib for organising a fine winter school. We thank
the SNAP weak lensing working group, especially Richard Ellis, for
our on-going collaboration and useful discussions.


%
\label{page:last}

\begin{references}

\reference Bacon, D., Refregier, A., Clowe, D.~\& Ellis, R.~2001,  
  \mnras{325}{1065}

\reference Bacon, D., Massey, R., Ellis, R.~\& Refregier, A.
  2003 {\it MNRAS} in press, preprint astro-ph/0203134

\reference Bartelmann, M.,~\& Schneider, P. 1999, astro-ph/9912508

\reference Benabed, K.,~\& Bernardeau, F.~2001, \prd{64}{083501}

\reference Benabed, K.,~\& van Waerbeke, L.~2003, astro-ph/0306033

\reference Bernardeau, F., van Waerbecke, L.~\& Mellier, Y.~1997,
  \aap{322}{1}

\reference Bernardeau, F., van Waerbecke, L.~\& Mellier, Y.~2002,
  \aap{389}{28}

\reference Clowe, D., Trentham, N.~\& Tonry, J.~2001,
  \aap{369}{16}

\reference Dahle, H.~\etal~2002, \apj{139}{313}

\reference Erben, T., van Waerbeke, L., Bertin, E., Mellier, Y.~\&
Schneider, P. 2001, \aap{366}{717}

\reference Jarvis, M.~\etal~2003, \aj{125}{1014}

\reference Jain, B., Seljak, U.~\& White, S.~2000, \apj{530}{547}

\reference Hamana, T.~\etal~2002, submitted to {\it ApJ}, preprint
  astro-ph/0210450

\reference Hoekstra H, Yee HKC, Gladders M. 2002b. astro-ph/0205205 

\reference Hu, W.~2001, \prd{66}{3515}  

\reference Hu, W.,~\& Tegmark, M.~1999, \apj{514}{L65}

\reference Hui, L.~1999, \apj{519}{9}

\reference Huterer, D.~2001, \prd{65}{063001}

\reference Huterer, D.~\& White, M.~2002, \apj{578}{L95}

\reference Kaiser, N., Squires, G.~\& Broadhurst, T.~1995, ApJ, 449, 460

\reference Kaiser, N., Tonry, J.~\& Luppino, G.~2000. {\it PASP} 112:768.
  {\it Pan-STARRS webpage} {\tt http://pan-starrs.ifa.hawaii.edu/}

\reference Lilly, S.~\etal~1996, \apj{455}{108}

\reference Ma, C.-P., Caldwell, R.R., Bode, P.,~\& Wang, L.~1999,
  \apj{521}{L1}

\reference Massey, R. \etal~2004a, {\it MNRAS} 348, 214

\reference Massey, R. \etal~2004b, {\it AJ} in press, preprint
  astro-ph/0304418

\reference Mellier, Y., 1999, \araa{37}{127}

\reference Mellier, Y.~\etal~2001, Cosmic shear surveys.  
  {\it Deep Fields, Proc. Eur. South. Obs.}, {Oct.}, {\it Garching}, 
  {Ger.} astro-ph/0101130

\reference Miyazaki, S.~\etal~2002, \apj{580}{L97}

\reference Munshi, D.,~\& Wang, Y.~2003, \apj{583}{566}

\reference Peacock, J.,~\& Dodds~1996, \mnras{280}{L19}

\reference Perlmutter, S.~\etal~1999, ApJ, 517, 565

\reference Perlmutter, S.~\etal~2003, {\it SNAP homepage}
  {\tt http://snap.lbl.gov}

\reference Pierpaoli, E., Scott, D.~\& White, M.~2001, \mnras{325}{77}

\reference Refregier, A.~2003a, \mnras{338}{35}

\reference Refregier, A.~2003b, \araa{41}{645}, astro-ph/0307212

\reference Refregier, A.~\& Bacon, D.~2003, \mnras{338}{48}

\reference Refregier, A.~\etal~2004, {\it AJ} in press, preprint
  astro-ph/0304419

\reference Rhodes, J., Refregier, A.~\& Groth, E.J.~2000,
  \apj{536}{79}

\reference Rhodes, J.~\etal~2004, Astropart. Phys. 20, 377

\reference Spergel, D.~\etal2003, submitted to {\it ApJ}, preprint
  astro-ph/0302209

\reference Taylor, A.~2001, Phys. Rev. Lett. submitted, preprint
  astro-ph/0111605

\reference Tyson, J., Wittman, D., Hennawi, J.~\& Spergel, D.~2002.  {\it
 Proc. 5th Int. UCLA Symp. Sources Detect. Dark Matter}, {Feb.}, {\it
 Marina del Rey}, ed. D Cline. astro-ph/0209632, LSST Home Page
 {\tt http://lsst.org}

\reference van Waerbeke, L., Hamana, T., Scoccimarro, R., Colombi, S.~\&  
 Bernardeau, F.~2001, \mnras{322}{918}
     
\reference van Waerbeke, L.~\etal~2002, submitted to {\it A\&A}, preprint
  astro-ph/0202503

\reference Viana, P.~\& Liddle, A.~1999, \mnras{303}{535}

\reference Weinberg, N., \& Kamionkowski, M., 2002,
  submitted to {\it MNRAS}, preprint astro-ph/0210134

\reference Williams, R.~\etal~1996, \aj{112}{1335}

\reference Williams, R.~\etal~1998, \aaps{193}{7501}

\reference Wittman DM. 2002. {\it Dark Matter and Gravitational
  Lensing}, {\it LNP Top. Vol.}, ed. F Courbin, D
  Minniti. Springer-Verlag. astro-ph/0208063

\end{references}
\end{document}